\newcommand{\CLASSINPUTinnersidemargin}{16.9mm}
\newcommand{\CLASSINPUTtoptextmargin}{19mm}
\newcommand{\CLASSINPUTbottomtextmargin}{25mm}
\documentclass[conference,a4paper]{IEEEtran}
\usepackage{ifpdf}
\usepackage[dvips]{graphicx}

\ifCLASSINFOpdf
\else
\fi
\usepackage[cmex10]{amsmath}
\usepackage{url}
\begin{document}
%
% paper title
% can use linebreaks \\ within to get better formatting as desired
\title{Cross-Correlation of Photovoltaic Output Fluctuation in Power System Operation for Large-Scale Photovoltaic Integration}
%\title{Cross-Correlation of Photovoltaic Output Fluctuation in Power System Operation toward Large-Scale Grid-Integration of Photovoltaic Systems}

% author names and affiliations
% use a multiple column layout for up to three different
% affiliations
\author{
\IEEEauthorblockN{Yuichi IKEDA}
\IEEEauthorblockA{Graduate School of Advanced \\Integrated Studies in Human Survivability, \\Kyoto University, Kyoto 606-8501, Japan}
\and
\IEEEauthorblockN{Kazuhiko OGIMOTO}
\IEEEauthorblockA{Institute of Industrial Science, \\University of Tokyo, Tokyo 153-8505, Japan}
}

% conference papers do not typically use \thanks and this command
% is locked out in conference mode. If really needed, such as for
% the acknowledgment of grants, issue a \IEEEoverridecommandlockouts
% after \documentclass

% for over three affiliations, or if they all won't fit within the width
% of the page, use this alternative format:
% 
%\author{\IEEEauthorblockN{Michael Shell\IEEEauthorrefmark{1},
%Homer Simpson\IEEEauthorrefmark{2},
%James Kirk\IEEEauthorrefmark{3}, 
%Montgomery Scott\IEEEauthorrefmark{3} and
%Eldon Tyrell\IEEEauthorrefmark{4}}
%\IEEEauthorblockA{\IEEEauthorrefmark{1}School of Electrical and Computer Engineering\\
%Georgia Institute of Technology,
%Atlanta, Georgia 30332--0250\\ Email: see http://www.michaelshell.org/contact.html}
%\IEEEauthorblockA{\IEEEauthorrefmark{2}Twentieth Century Fox, Springfield, USA\\
%Email: homer@thesimpsons.com}
%\IEEEauthorblockA{\IEEEauthorrefmark{3}Starfleet Academy, San Francisco, California 96678-2391\\
%Telephone: (800) 555--1212, Fax: (888) 555--1212}
%\IEEEauthorblockA{\IEEEauthorrefmark{4}Tyrell Inc., 123 Replicant Street, Los Angeles, California 90210--4321}}

% use for special paper notices
%\IEEEspecialpapernotice{(Invited Paper)}

% make the title area
\maketitle

\begin{abstract}
%\boldmath
We analyzed the cross-correlation of Photovoltaic (PV) output fluctuation for the actual PV output time series data in both the Tokyo area and the whole of Japan using the principal component analysis with the random matrix theory. Based on the obtained cross-correlation coefficients, the forecast error for PV output was estimated with/without considering the cross-correlations. Then operation schedule of thermal plants is calculated to integrate PV output using our unit commitment model with the estimated forecast error. The cost for grid integration of PV system was also estimated. Finally, validity of the concept of ``local production for local consumption of renewable energy" and alternative policy implications were also discussed.
\end{abstract}
% IEEEtran.cls defaults to using nonbold math in the Abstract.
% This preserves the distinction between vectors and scalars. However,
% if the conference you are submitting to favors bold math in the abstract,
% then you can use LaTeX's standard command \boldmath at the very start
% of the abstract to achieve this. Many IEEE journals/conferences frown on
% math in the abstract anyway.

% no keywords

% For peer review papers, you can put extra information on the cover
% page as needed:
% \ifCLASSOPTIONpeerreview
% \begin{center} \bfseries EDICS Category: 3-BBND \end{center}
% \fi
%
% For peerreview papers, this IEEEtran command inserts a page break and
% creates the second title. It will be ignored for other modes.
\IEEEpeerreviewmaketitle

\section{Introduction}\label{sec:intro}

Restructuring of the electric utility industry and large-scale grid-integration of PV systems are intensively discussed after the East Japan Earthquake of 2011. The former includes separation of electrical power generation from power distribution and transmission, and establishment of the retail power market and revitalization of the wholesale power markets. Although the institutional design of the power markets is still an open question in Japan, the market has to be designed so as to have optimal operation schedule, which is obtained using a unit commitment calculation, through competitions between generation companies. 

The large-scale grid-integration of PV systems brings another kind of problem, namely, the PV output fluctuation, into the power system operation. The planed installation capacity of PV systems will be 100 GW and 33 GW in 2030 in the whole of Japan and the Tokyo area, respectively \cite{PVoutlook2030}. The major fraction of the PV system will be installed on the rooftop of the consumer's residential houses and office buildings, which are widely distributed in the Tokyo area. Therefore, the forecast of PV output with high spatial resolution is an important problem to be considered, and the cross-correlations of the PV outputs will be key quantities to estimate the forecast error of PV output.

In relation to the above discussion, the concept of ``local production for local consumption of renewable energy" has been proposed in Japan. Because electric power is in large demand in the Tokyo area, the area price could be high enough to be close to the feed-in tariff price for PV power. For this reason, the concept of ``local production for local consumption of renewable energy" of PV power is considered to be economically feasible \cite{Hatta2012}. This concept is also advantageous because of the mitigation of transmission loss. However, it is to be noted that this concept needs careful consideration for PV and wind power because of the inherent nature of output fluctuation, even though it is suitable for geothermal and biomass energies \cite{Niitsuma2003}.

In this paper, we analyzed the cross-correlation of PV output fluctuation for the actual PV output time series data \cite{Ozeki2011} in both Tokyo area and the whole of Japan using the principal component analysis with the random matrix theory. Based on the obtained cross-correlation coefficients, the forecast error for PV output was estimated for some extreme cases. Then the operation schedule of thermal plants was calculated to integrate PV output using our unit commitment model  \cite{Ikeda2012, Ikeda2013} with the estimated forecast error. The cost for grid integration of PV system was also estimated. Finally, validity of the concept of ``local production for local consumption of renewable energy" and alternative policy implications were also discussed.

\section{Cross-Correlation of PV Output Fluctuation}\label{sec:cross}

\subsection{System-Wide Output Fluctuation}\label{sec:cross:fluct}

The forecast of system-wide PV output is decomposed as
\begin{equation}
 \label{eq:cross1}
 pv_t^{(f)} \equiv X(t) = \sum_{i=1}^N x_i(t) = \sum_{i=1}^N c_i y_i(t),
\end{equation}
where $y_i(t)=x_i(t)/c_i$ and $c_i$  are the forecast of PV output per installed capacity (load factor) and the installed capacity in the i-th site, respectively. Our unit commitment model \cite{Ikeda2012, Ikeda2013} requires the PV output forecast time-series and the forecast error to estimate the optimal operation schedule with consideration of the PV output fluctuation. 
If both accuracy and spatial resolution of the PV forecasting is high, the forecasted time-series is similar to a moving average of actual PV output for each PV site, and consequently the cross-correlation of residual time-series, which is equal to subtracting the actual output from forecast output at each time point, is expected to be a white noise. Thus, the forecast error of system-wide PV output $\sigma_X$  is
\begin{equation}
 \label{eq:cross2}
 \sigma_p^2 = \sum_{i=1}^N \Bigl( \frac{\partial X}{\partial y_i} \Bigr)^2 \sigma_i^2 = \sum_{i=1}^N c_i^2 \sigma_i^2,
\end{equation}
where $\sigma_i$  is the forecast error of PV output per installed capacity in the i-th site.
 On the other hand, if the spatial resolution of the forecast is low and, for example, we have just a few forecasted sites in the Tokyo area, the residual time-series includes the cross-correlation between the various PV sites located in different places. In this case, we have a larger forecasting error due to the cross-correlations. The forecast error of system-wide PV output $\sigma_X$  is written as
\begin{equation}
\begin{split}
 \label{eq:cross3}
\sigma_p^2 &= \sum_{i=1}^N \Bigl( \frac{\partial X}{\partial y_i} \Bigr)^2 \sigma_i^2 + 2 \sum_{i=2}^N \sum_{j<i} \Bigl( \frac{\partial X}{\partial y_i} \Bigr) \Bigl( \frac{\partial X}{\partial y_j} \Bigr) \sigma_{ij} \\
&= \sum_{i=1}^N c_i^2 \sigma_i^2  + 2 \sum_{i=2}^N \sum_{j<i} c_i c_j \sigma_{ij} 
\end{split}
\end{equation}
\begin{equation}
 \label{eq:cross4}
\sigma_{ij} = \sigma_i \sigma_j \rho_{ij}
\end{equation}
by including covariance among different sites $\sigma_{ij}$. Here, $\rho_{ij}$ is the cross-correlation coefficient among different sites. Generally, it is expected that the number of forecasted sites is smaller than that of the installed sites $N$. For instance, we cannot forecast PV output for each roof-top PV of all the residential houses and office buildings with high accuracy in the Tokyo area due to both technological and economical reasons. Therefore, it is required to consider the cross-correlation $\sigma_{ij}$ to estimate the forecast error of system-wide PV output $\sigma_X$. 

\begin{figure}
\begin{center}
\includegraphics[width=0.3\textwidth]{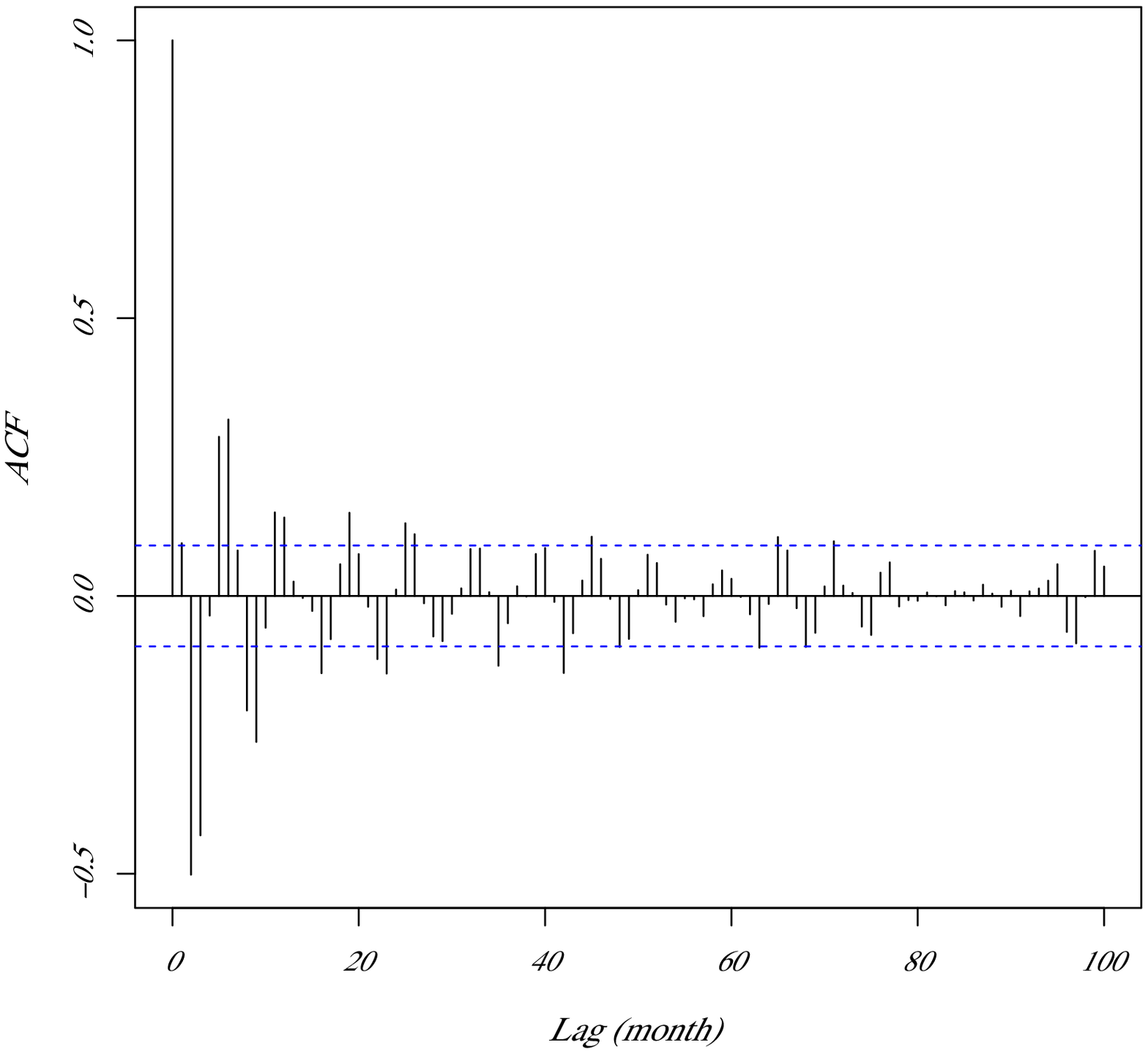}
\includegraphics[width=0.3\textwidth]{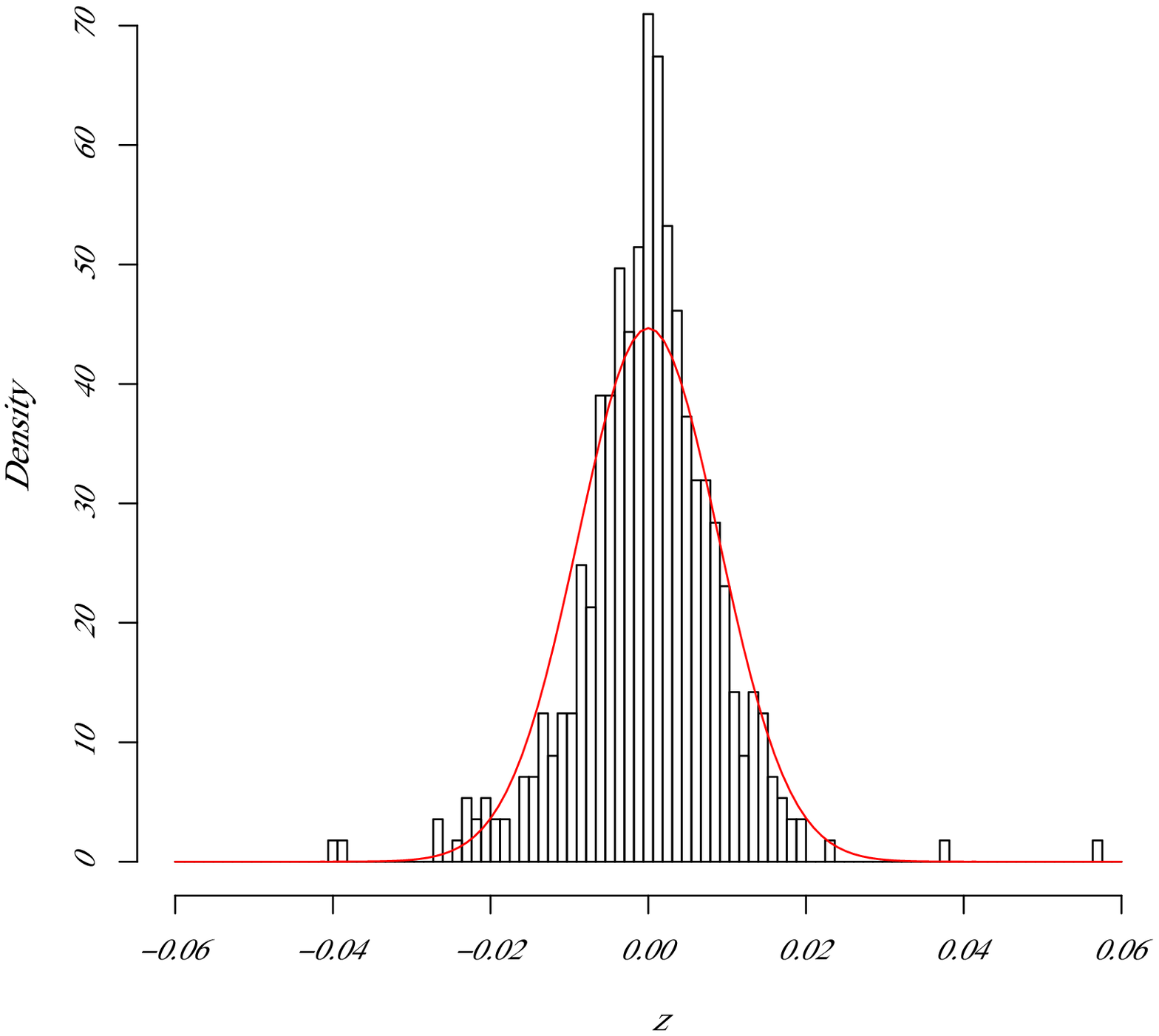}
\caption{
Auto-correlation function (top) and fluctuation distribution (bottom) for Tokyo in May
}
\label{fig:BasicTokyo}
\end{center}
\end{figure}
\begin{figure}
\begin{center}
\includegraphics[scale=0.5]{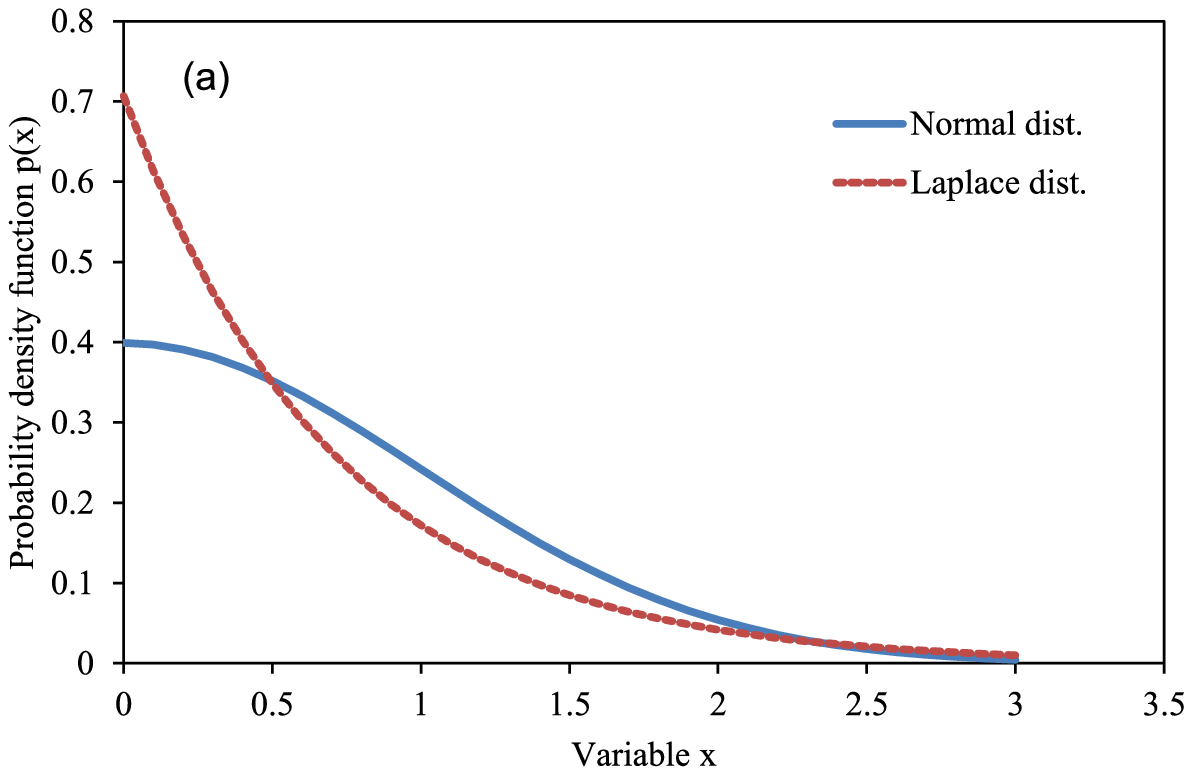}
\includegraphics[scale=0.5]{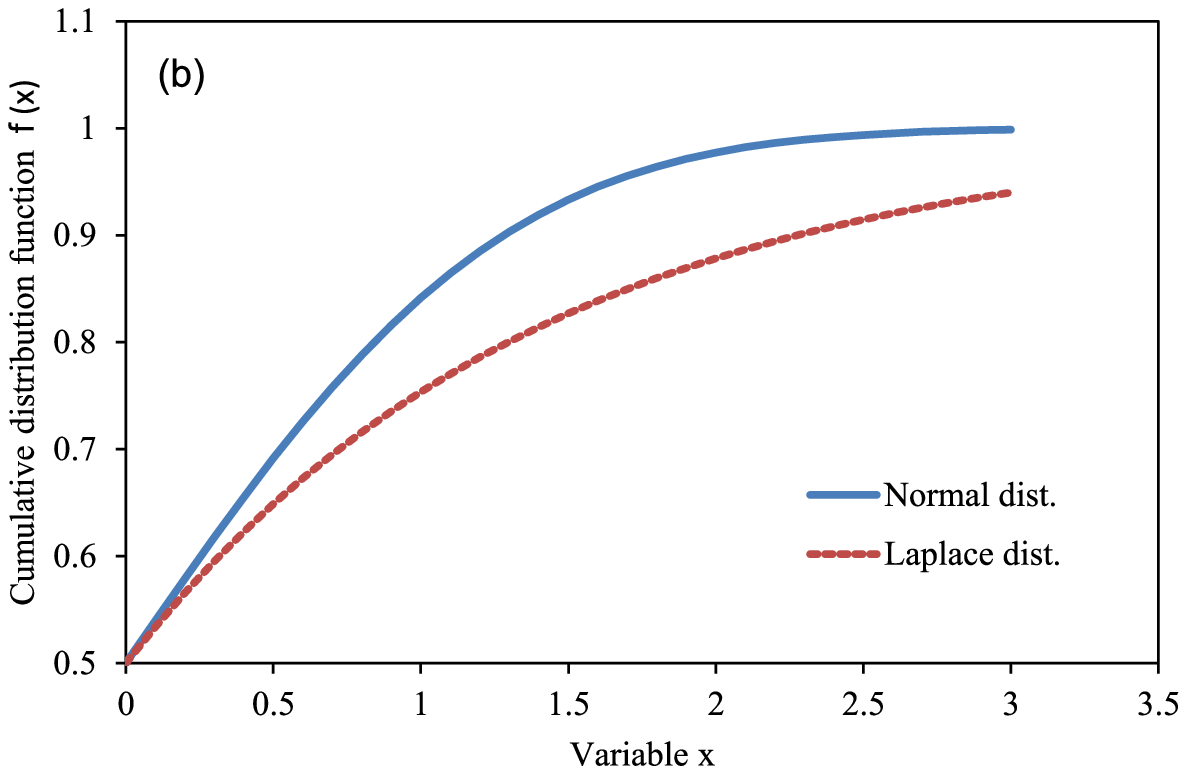}
\caption{Functional Form of Fluctuation Distribution}
\label{fig:dist}
\end{center}
\end{figure}
\begin{figure}
\begin{center}
\includegraphics[width=0.3\textwidth]{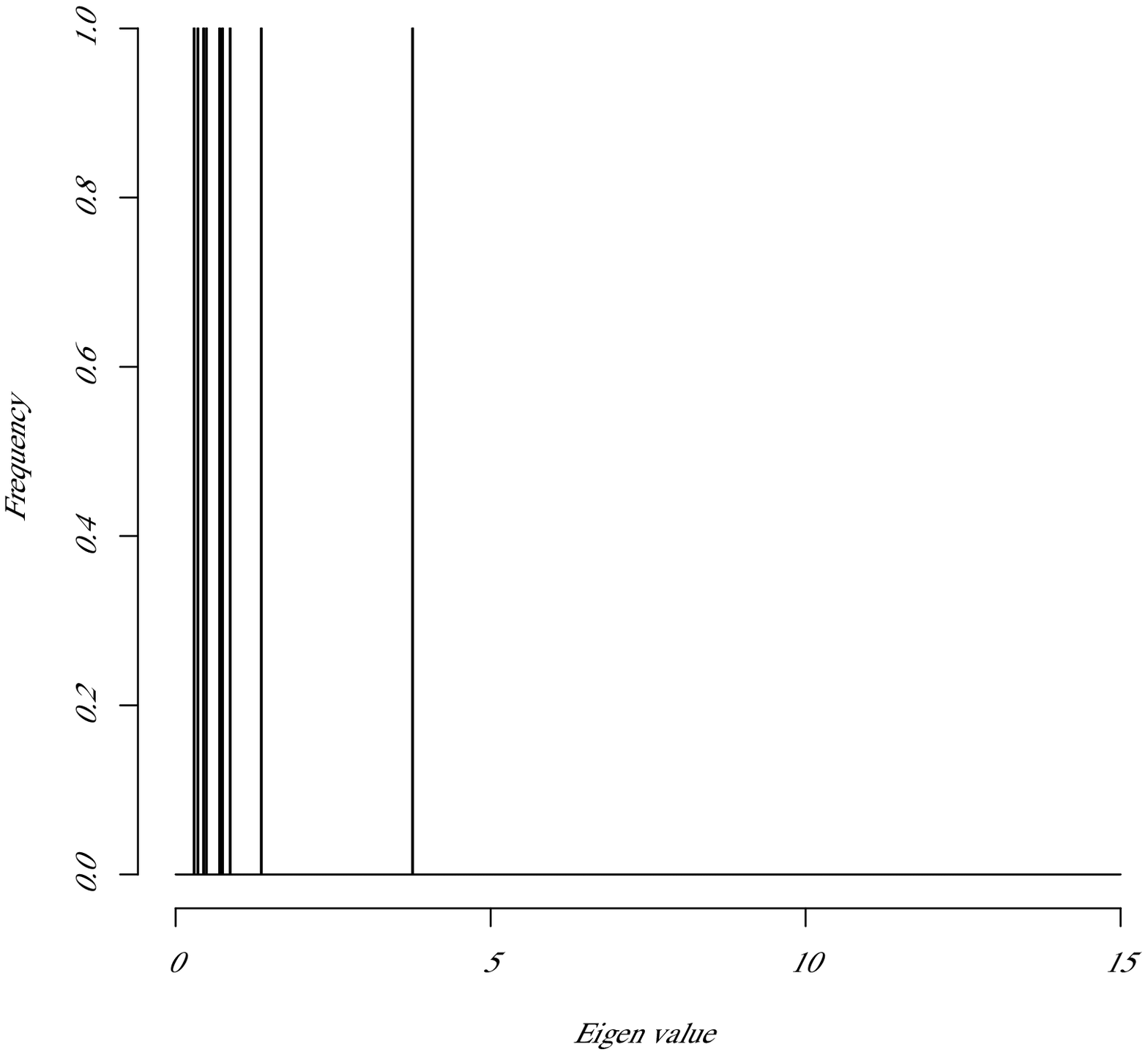}
\includegraphics[width=0.3\textwidth]{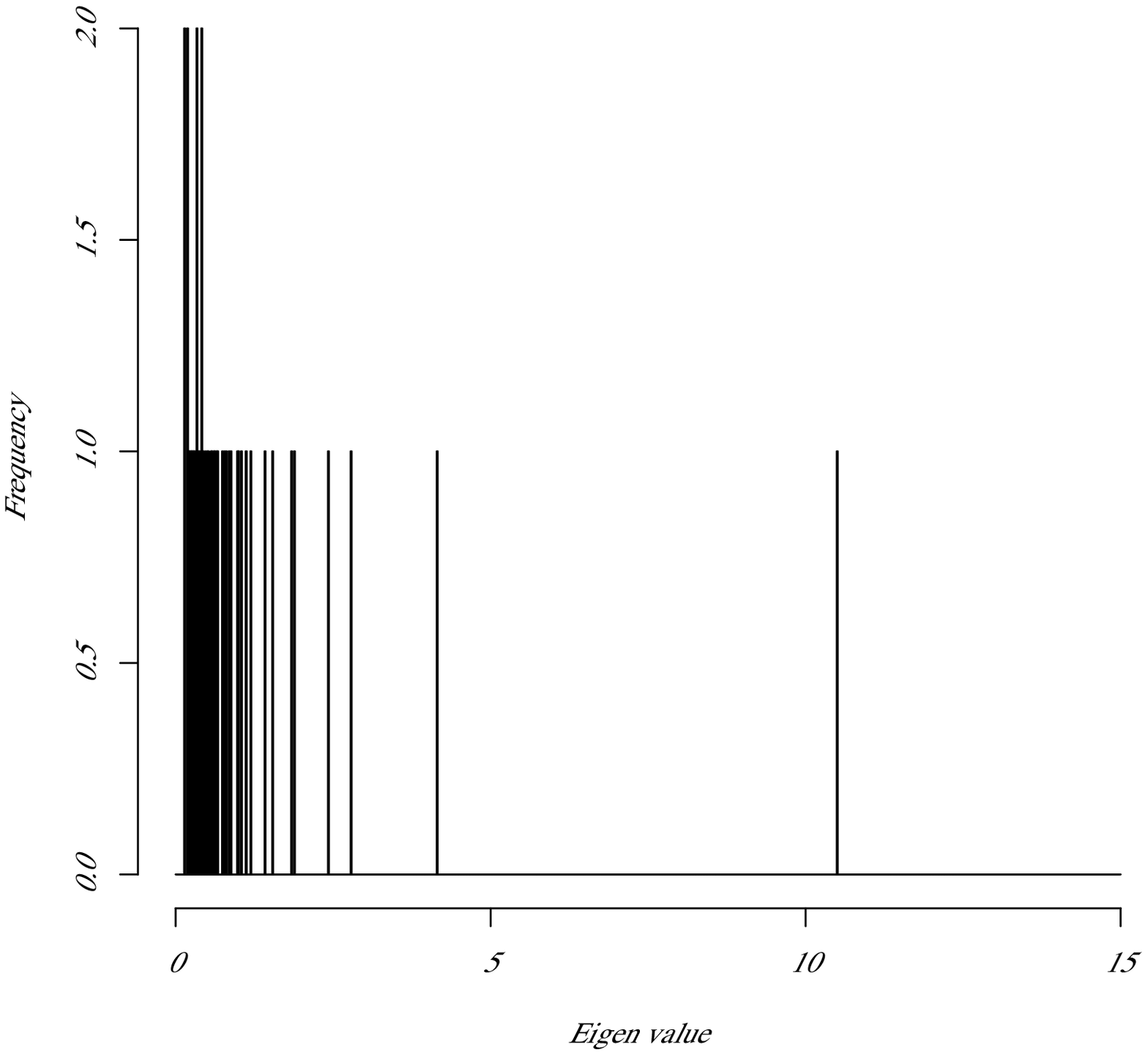}
\caption{
Eigen value distribution for the Tokyo area (top) and whole of Japan (bottom) in May
}
\label{fig:Eigen}
\end{center}
\end{figure}

\subsection{Random Matrix Theory}\label{sec:cross:rmt}

We analyzed the de-trended PV output $z_i(t)$ obtained by filtering the actual PV output time-series per installed capacity $y_i(t)$ using the Fourier series expansion.
In general it is expected that correlation coefficients are associated with random noise for a fluctuating time series such as PV output. The correlation coefficient between points $i$ and $j$ is calculated by

\begin{equation} 
C_{ij} = \frac{\langle (z_i(t)-\langle z_i \rangle) (z_j(t)-\langle z_j \rangle) \rangle}{\sqrt{(\langle z_i^2 \rangle - \langle z_i \rangle^2)(\langle z_j^2 \rangle - \langle z_j \rangle^2)}} 
\label{eq:CorrCoef} 
\end{equation}%
where $z_i(t)$ is the de-trended PV output at the site $i (=1,\cdots,N)$ and time $t (=1,\cdots,L)$ and $\langle \cdot \rangle$ indicates the time average for the time series.

Now we consider the eigen-value problem 
\begin{equation} 
C | \alpha \rangle = \lambda_\alpha | \alpha \rangle
\label{eq:Eigen}
\end{equation}%
for the correlation matrix $C$. $\lambda_\alpha$ and $| \alpha \rangle$ are the eigen-value and the corresponding eigen-vector, respectively.
Here, we assume that the eigen-values are arranged in decreasing order $(\alpha=0,\cdots,N-1)$.
Once the eigen-values are calculated using Eqs. (\ref{eq:CorrCoef}) and (\ref{eq:Eigen}),
the distribution of eigen-value $\rho ( \lambda )^E$ is obtained.

According to the random matrix theory \cite{Metha1991,Laloux1999,Pleroux1999,Pleroux2002}, distribution of the eigen-value for the matrix $\frac{1}{T} H H^T$ where all elements of the matrix $H$ are given as a random number $N(0,\sigma^2)$ is given by
\begin{equation} 
\rho ( \lambda )^T = \frac{Q}{2 \pi} \frac{\sqrt{(\lambda_{max}-\lambda)(\lambda-\lambda_{min})}}{\lambda},
\label{eq:RMT1}
\end{equation}%
where
\begin{equation}
Q=\frac{L}{N}, 
\label{eq:RMT2}
\end{equation}%
\begin{equation}
\lambda=[\lambda_{min},\lambda_{max}],
\label{eq:RMT3}
\end{equation}%
\begin{equation}
\lambda_{min}=(1-\frac{1}{\sqrt{Q}})^2, and
\label{eq:RMT4}
\end{equation}%
\begin{equation}
\lambda_{max}=(1+\frac{1}{\sqrt{Q}})^2.
\label{eq:RMT5}
\end{equation}%

Eq. (\ref{eq:RMT1}) is exact at the limit $N,L \to \infty$.
For a randomly fluctuating time series such as PV output, 
it is expected that the distribution $\rho(\lambda)^E$ obtained by data analysis agrees to the distribution $\rho(\lambda)^T$ calculated 
using Eqs. (\ref{eq:RMT1}) to (\ref{eq:RMT5}) for $\lambda \leq \lambda_{max}$. 
Therefore only the small number of eigen-values for $\lambda > \lambda_{max}$ have the information of genuine correlation.

In order to extract the genuine correlation, we rewrite the correlation matrix $C$ using eigen-value $\lambda_\alpha$ and the corresponding eigen-vector $| \alpha \rangle$ \cite{Iyetomi2011}.
First we define the complex conjugate vector of the eigen-vector $| \alpha \rangle$ by
\begin{equation}
\langle \alpha | = | \alpha^* \rangle ^t.
\label{eq:ccAlpha}
\end{equation}%
For the real symmetric matrix, such as the correlation matrix $C$, all elements of the eigen-vector $| \alpha \rangle$ are real 
and thus the complex conjugate denotes the transpose $t$.

Then the correlation matrix $C$ is rewritten as
\begin{equation}
C = \textstyle\sum\limits_{\alpha=0}^{N-1} \lambda _\alpha | \alpha \rangle \langle \alpha | 
\label{eq:NewCorrCoef}
\end{equation}%
by multiplying Eq. (\ref{eq:Eigen}) with the transposed vector $\langle \alpha |$ from the left hand side and taking summation over $\alpha$.
Here, the property of the projection operator $| \alpha \rangle \langle \alpha |$
\begin{equation}
\textstyle\sum\limits_{\alpha=0}^{N-1}| \alpha \rangle \langle \alpha | = 1
\label{eq:PrjOpr}
\end{equation}%
was used. As a result, the correlation matrix $C$ of Eq. (\ref{eq:NewCorrCoef}) is divided in the following components:
\begin{equation}
C = C^t + C^r = \textstyle\sum\limits_{\alpha=0}^{N_t} \lambda_\alpha | \alpha \rangle \langle \alpha | + \textstyle\sum\limits_{\alpha=N_t+1}^{N-1} \lambda _\alpha | \alpha \rangle \langle \alpha |. 
\label{eq:NewCorrCoef2}
\end{equation}%
The first term $C^t$ corresponds to the genuine correlation component ($\lambda > \lambda_{max}$).
The second term $C^r$ corresponds to the random component ($\lambda \leq \lambda_{max}$).
The term $\lambda_0 | 0 \rangle \langle 0 |$ is interpreted as the change as a whole system, such as the weather change.

We introduce the vector $| z(t) \rangle$, which consists of the time series of PV output $z_i(t) (i=1,\cdots,N)$.
Then the vector $| z(t) \rangle$ is expanded on the basis of the eigen-vectors $| \alpha \rangle$ \cite{Iyetomi2011}
:
\begin{equation}
| z(t) \rangle =  \textstyle\sum\limits_{\alpha=0}^{N-1} a_{\alpha}(t) | \alpha \rangle.
\label{eq:VctExp}
\end{equation}%
The expansion coefficient $a_{\alpha}(t)$ is obtained using the orthogonality of the eigen-vectors:
\begin{equation}
a_{\alpha}(t) = \langle \alpha | z(t) \rangle.
\label{eq:ExpCoef}
\end{equation}%
The time series corresponding to the genuine correlation $C^t$ is extracted by truncating the summation up to $N_t$ in Eq.(\ref{eq:VctExp}):
\begin{equation}
| z(t) \rangle =  \textstyle\sum\limits_{\alpha=0}^{N_t} a_{\alpha}(t) | \alpha \rangle.
\label{eq:VctExp2}
\end{equation}%

\subsection{Data analysis}

The genuine components of cross-correlation of the de-trended PV output per installed capacity were studied using the random matrix theory. 
The analyzed data is the output time series acquired every one hour for each prefecture \cite{Ozeki2011}. 
Before analyzing the data, two preprocessing were made. First, the data during night time was removed. Then, the trend was removed from the time series by filtering out the components with a period longer than six hours using the Fourier series expansion.
Therefore, only the short-term fluctuation is component remained in the time series. 
The auto-correlation function and fluctuation distribution for Tokyo in May are shown in Fig. \ref{fig:BasicTokyo}.
The memory in the auto-correlation function gets lost within a few hours. This means that the trend component is well removed. 
The kurtosis of the fluctuation distribution is 5.0849, which is significantly larger than the value expected for the normal distribution, i.e. 3.0.
This means that the actual fluctuation distribution has a longer tail than the normal distribution.
The two different types of functional forms of fluctuation distribution are shown in Fig. \ref{fig:dist}.
If the fluctuation is distributed according to the normal distribution, the probability density function is
\begin{equation}
p(x) = \frac{1}{\sqrt{2 \pi \sigma^2}} exp \Big[-\frac{(x-\mu)^2}{2 \sigma^2} \Big],
\label{eq:NormalPDF}
\end{equation}
and, the cumulative distribution function is written using the error function $erf[\cdot]$ as 
\begin{equation}
\phi(x) = \frac{1}{2} \Big( 1 + erf \Big[\frac{x-\mu}{\sqrt{2 \sigma^2}} \Big] \Big),
\label{eq:NormalCDF}
\end{equation}
where $\mu$ and $\sigma$ are the mean and standard deviation, respectively.
However, if the probability density function $p(x)$ is a Laplace distribution
\begin{equation}
p(x) = \frac{1}{2b} exp \Big[-\frac{|x-\mu|}{b} \Big],
\label{eq:LaplacePDF}
\end{equation}
then, the cumulative distribution function $\phi(x)$ is
\begin{equation}
\phi(x) = \frac{1}{2} \Big( 1 + sgn(x-\mu) \Big(1 - exp \Big[-\frac{|x-\mu|}{b} \Big] \Big) \Big).
\label{eq:LaplaceCDF}
\end{equation}
Here, a standard deviation is given by $\sigma=\sqrt{2}b$ and $sgn(x-\mu)=+(x\ge\mu), -(x<\mu)$.
The functional forms for these distributions are depicted for $\mu=0$ and $\sigma=1$ in Fig. \ref{fig:dist}.
It is to be noted here that the Laplace distribution shows a distribution tail longer than the normal distribution.

Eigen-value distribution for the Tokyo area and the whole of Japan in May is shown in Fig. \ref{fig:Eigen}.
For the Tokyo area, we calculate $\lambda_{max}=1.35$ using Eq. (\ref{eq:RMT5}) with $N=9$ and $L=420$.
The upper panel of Fig. \ref{fig:Eigen} depicts that only the largest eigen-value is larger than $\lambda_{max}$.
On the other hand, for the whole of Japan, we calculate $\lambda_{max}=1.88$ with $N=47$ and $L=420$.
The lower panel of Fig. \ref{fig:Eigen} depicts that the five largest eigen-values are larger than $\lambda_{max}$.

\begin{figure}
\begin{center}
\includegraphics[width=0.4\textwidth]{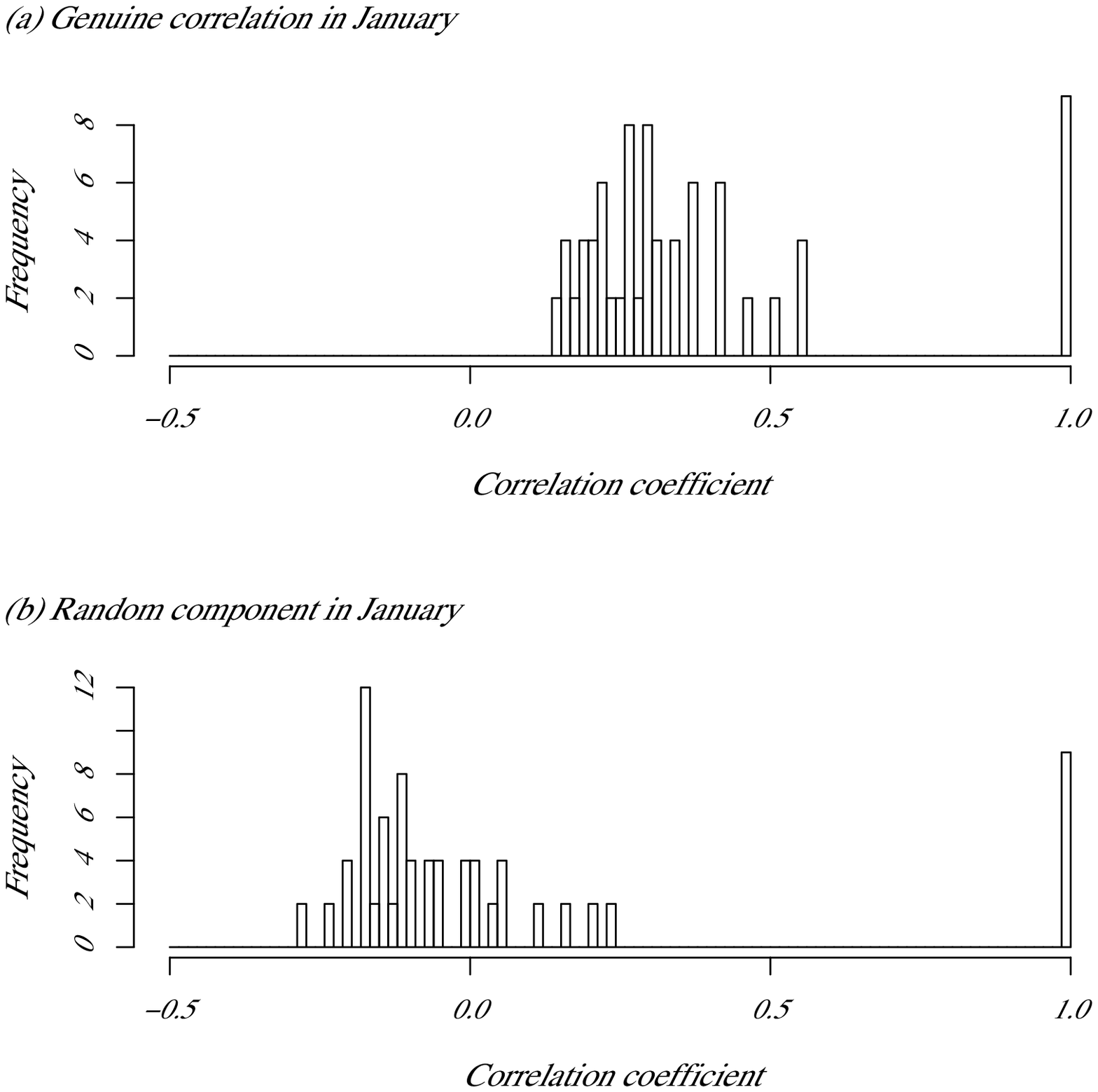}
\includegraphics[width=0.4\textwidth]{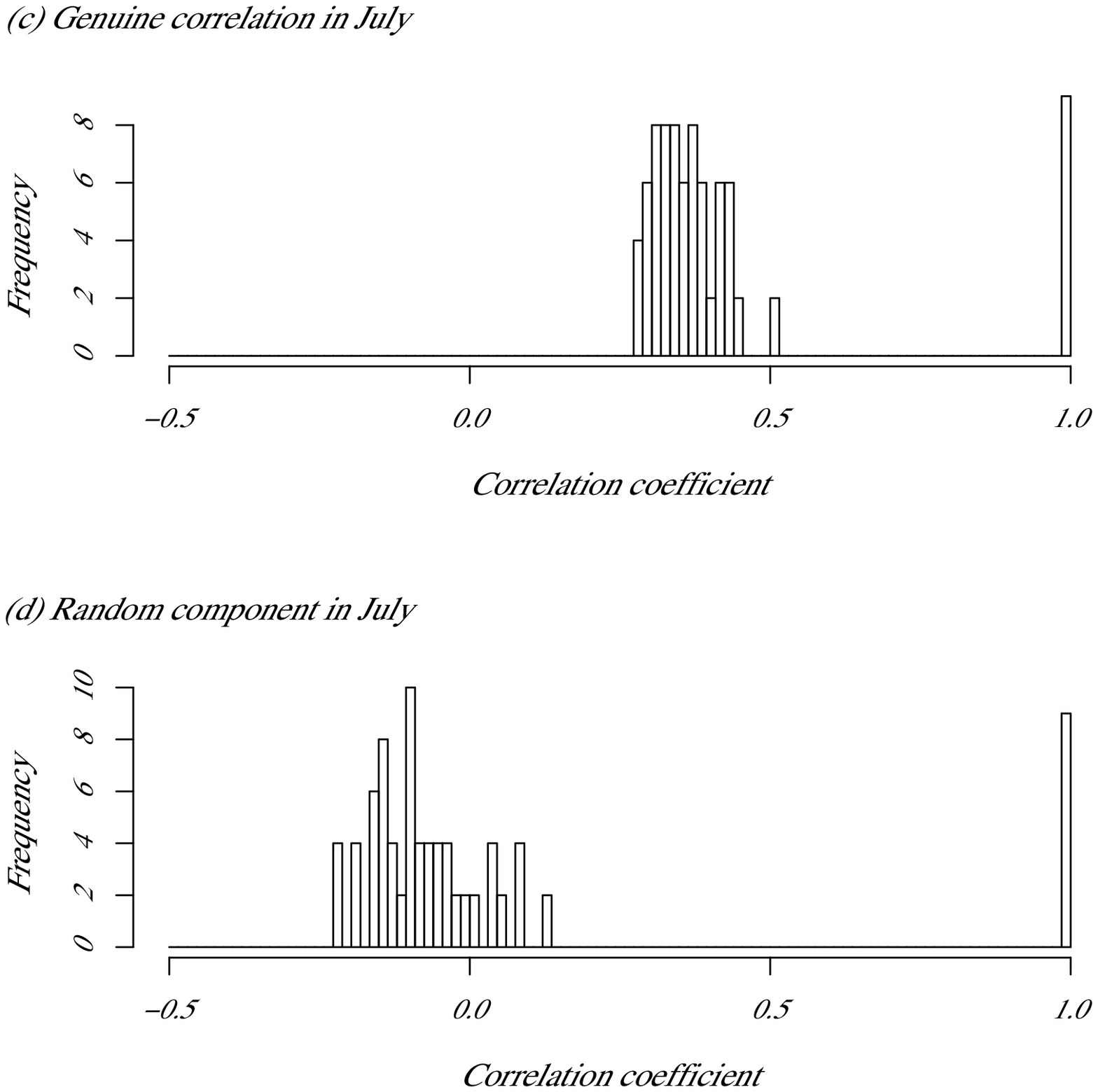}
\caption{
Cross-correlation coefficients for the Tokyo area in January and July
}
\label{fig:CrossCorrTEPCO}
\end{center}
\end{figure}

\begin{figure}
\begin{center}
\includegraphics[width=0.4\textwidth]{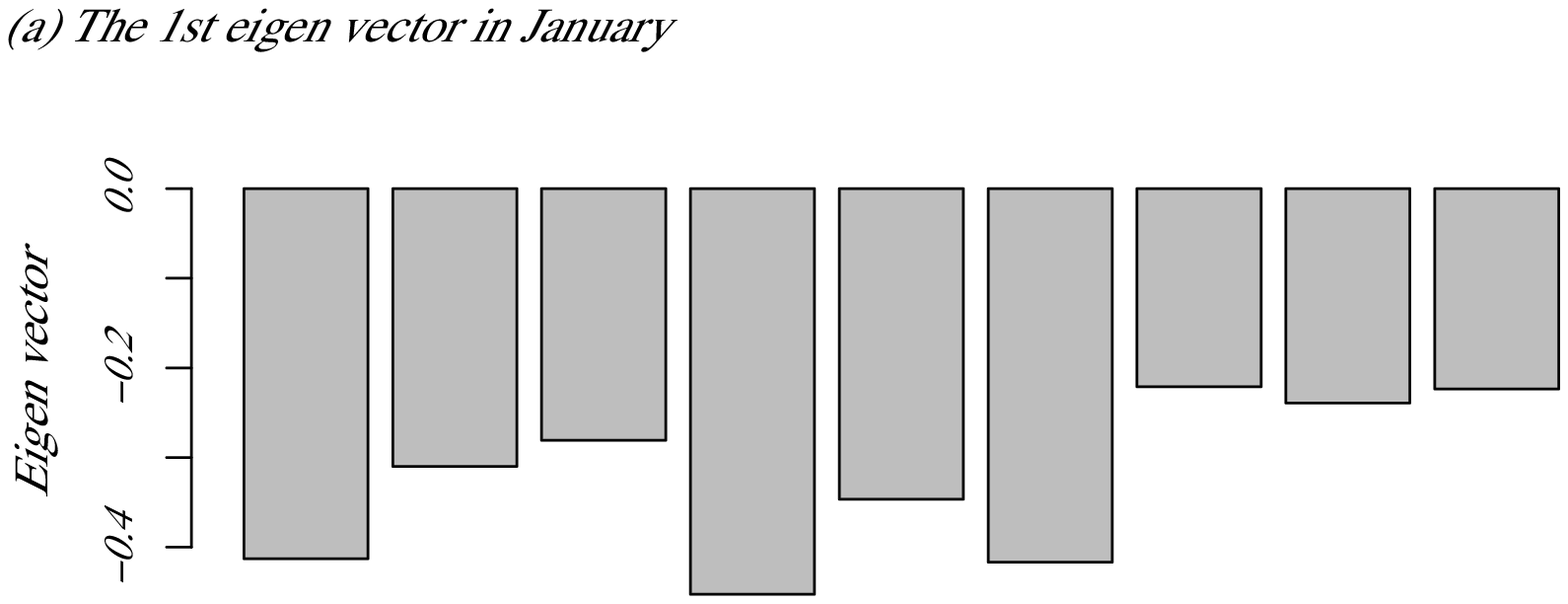}
\includegraphics[width=0.4\textwidth]{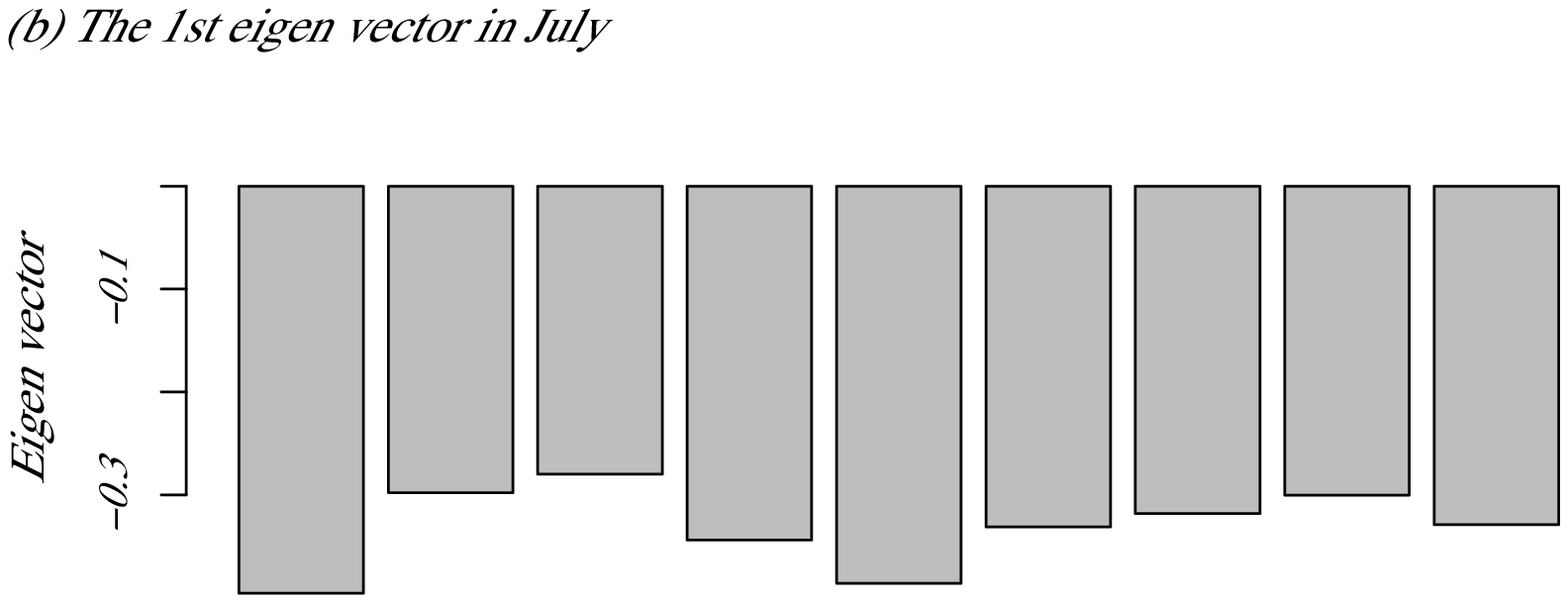}
\caption{
The 1st eigen-vector for the Tokyo area in January and July
}
\label{fig:EigenVectorTEPCO}
\end{center}
\end{figure}

We show the distribution of genuine correlation coefficients calculated for the de-trended PV output time series in both the Tokyo area and the whole of Japan.
The cross-correlation coefficients for the Tokyo area are shown in Fig. \ref{fig:CrossCorrTEPCO}.
Panels (a) and (c) are genuine correlation $C^t$ and panels (b) and (d) are the random components $C^r$.
The genuine correlation $C^t$ was calculated using only the largest eigen-value and the corresponding eigen-vector.
Figure \ref{fig:CrossCorrTEPCO} depicts that the genuine correlation $C^t$ has positive correlation and on the other hand the random components $C^r$ distributes around $0.0$.
The 1st eigen-vector for the Tokyo area is shown in Fig. \ref{fig:EigenVectorTEPCO}. 
The nine components correspond to eight prefectures and Tokyo was included in the Tokyo area.
It was noted that all vector components had the same sign. This means that the PV output fluctuates simultaneously in the same direction for all prefectures in the Tokyo area.

\begin{figure}
\begin{center}
\includegraphics[width=0.4\textwidth]{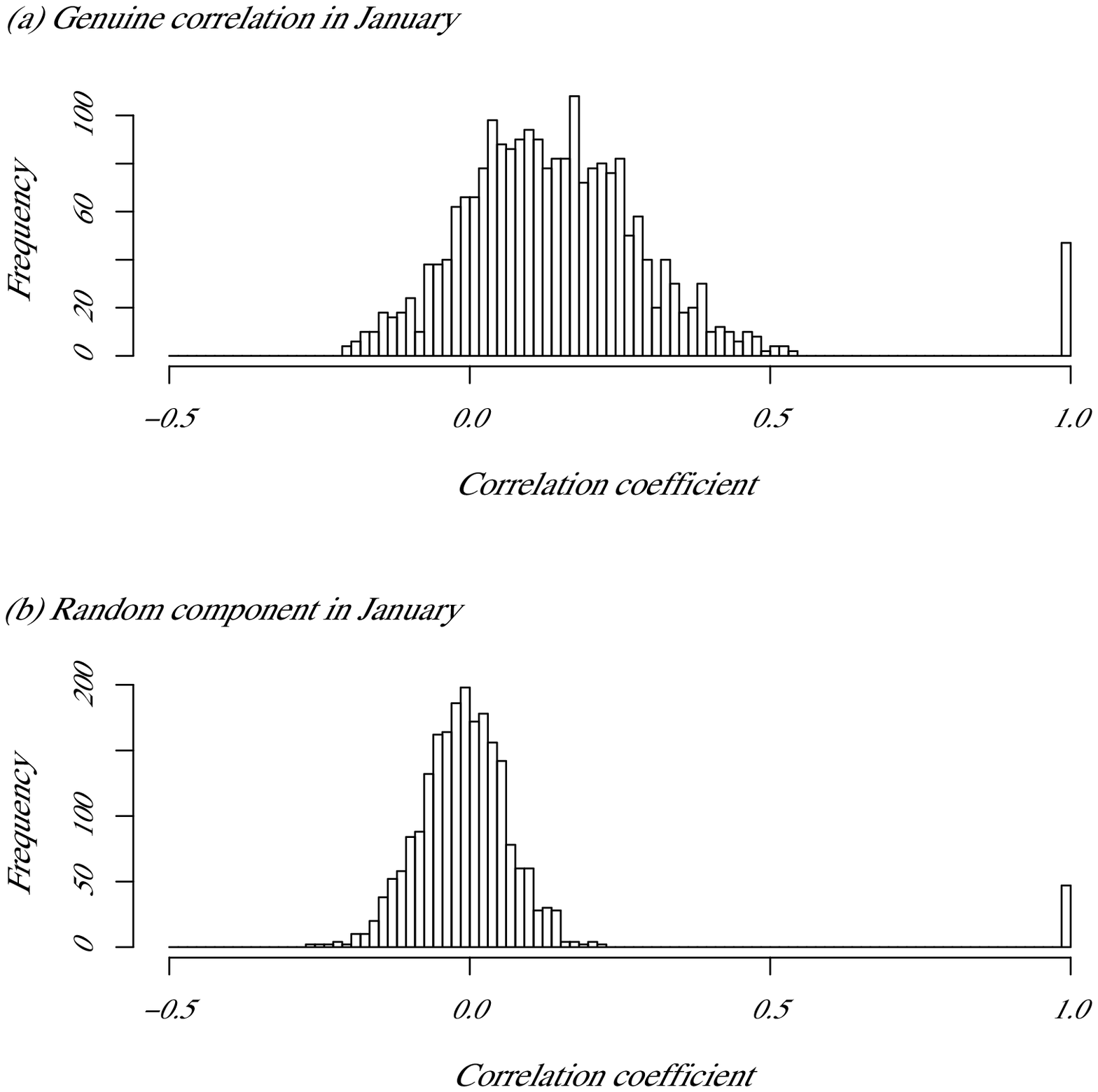}
\includegraphics[width=0.4\textwidth]{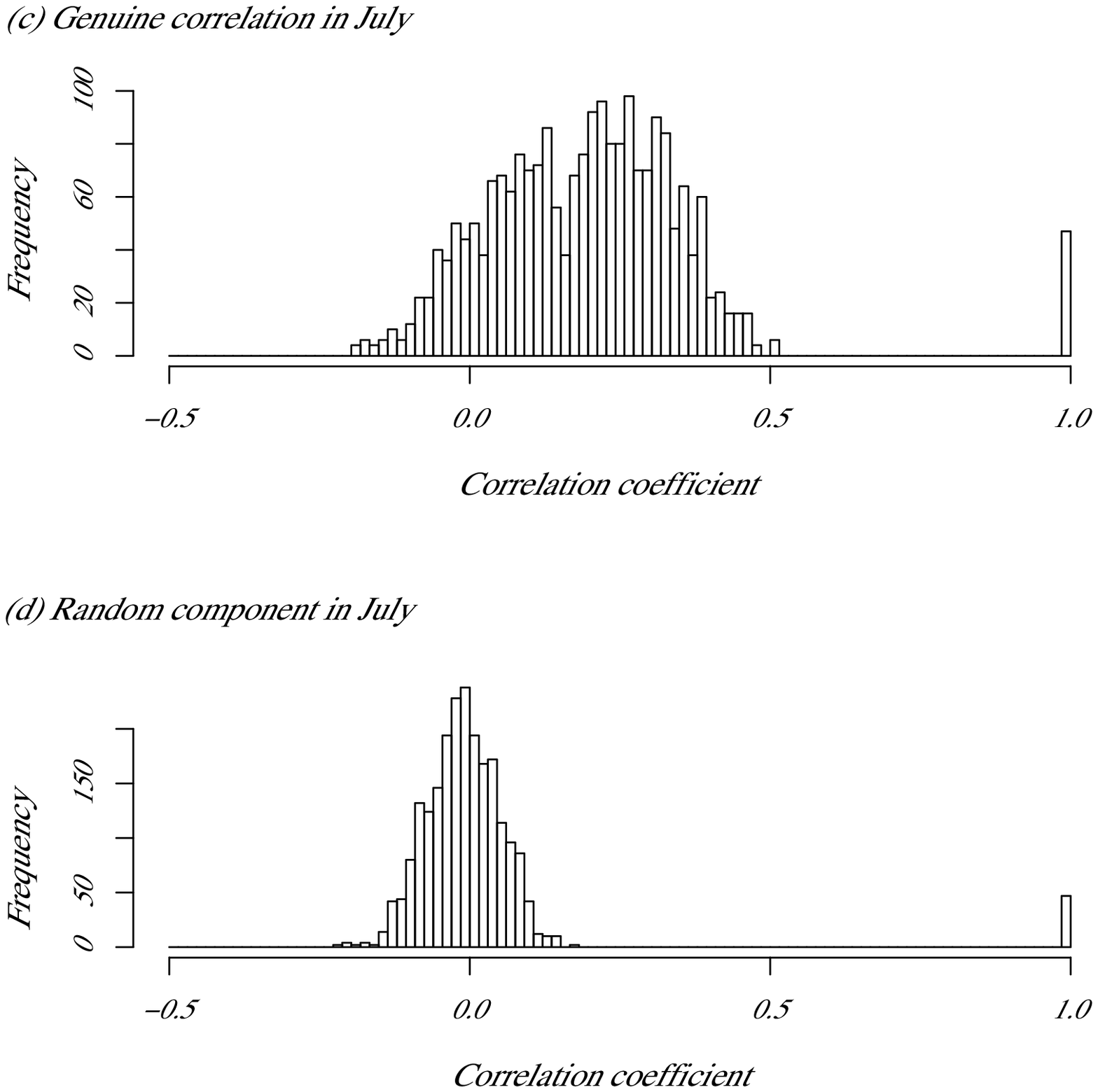}
\caption{
Cross-correlation coefficients for the whole of Japan in January and July
}
\label{fig:CrossCorrJapan}
\end{center}
\end{figure}

\begin{figure}
\begin{center}
\includegraphics[width=0.4\textwidth]{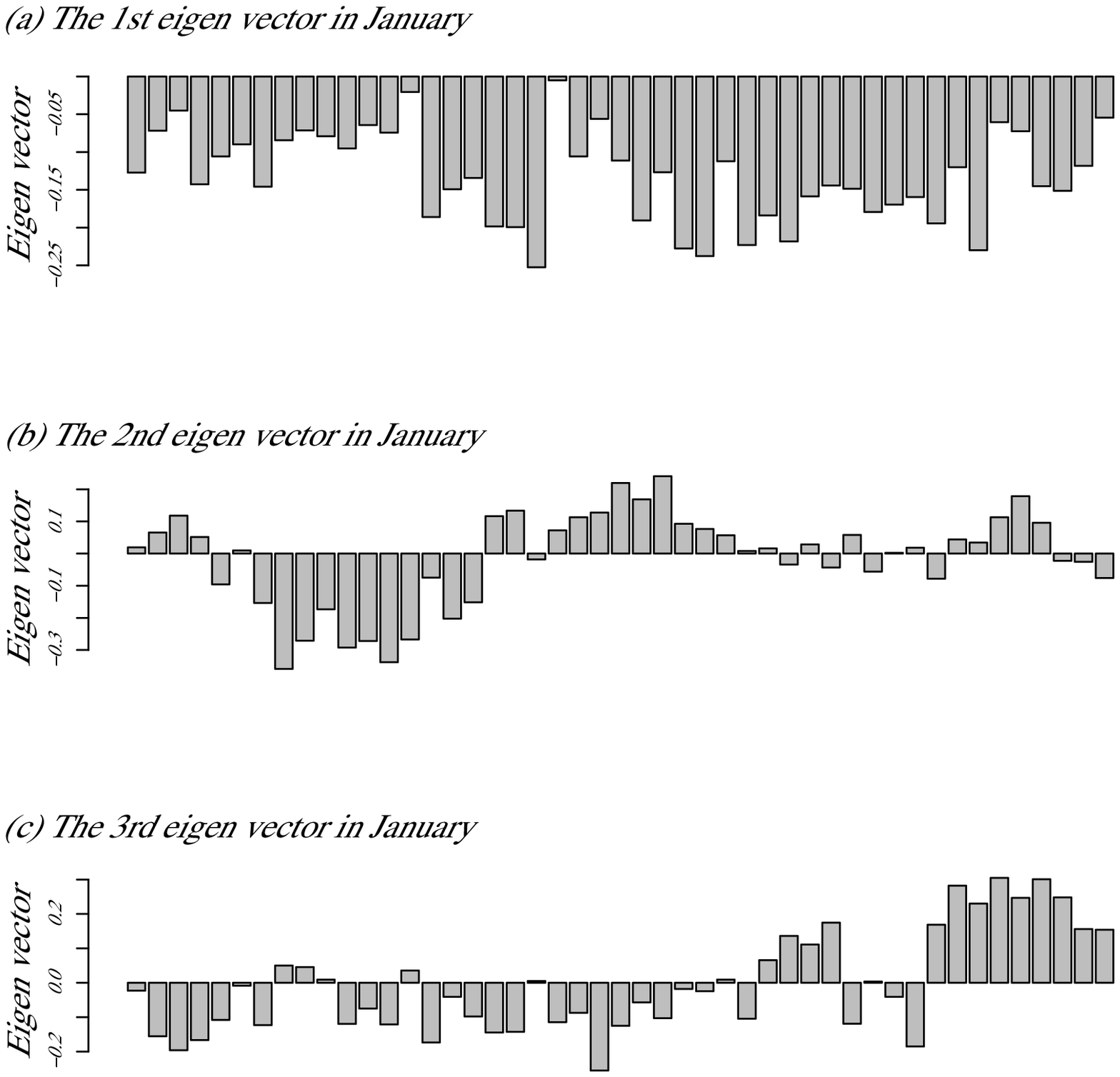}
\includegraphics[width=0.4\textwidth]{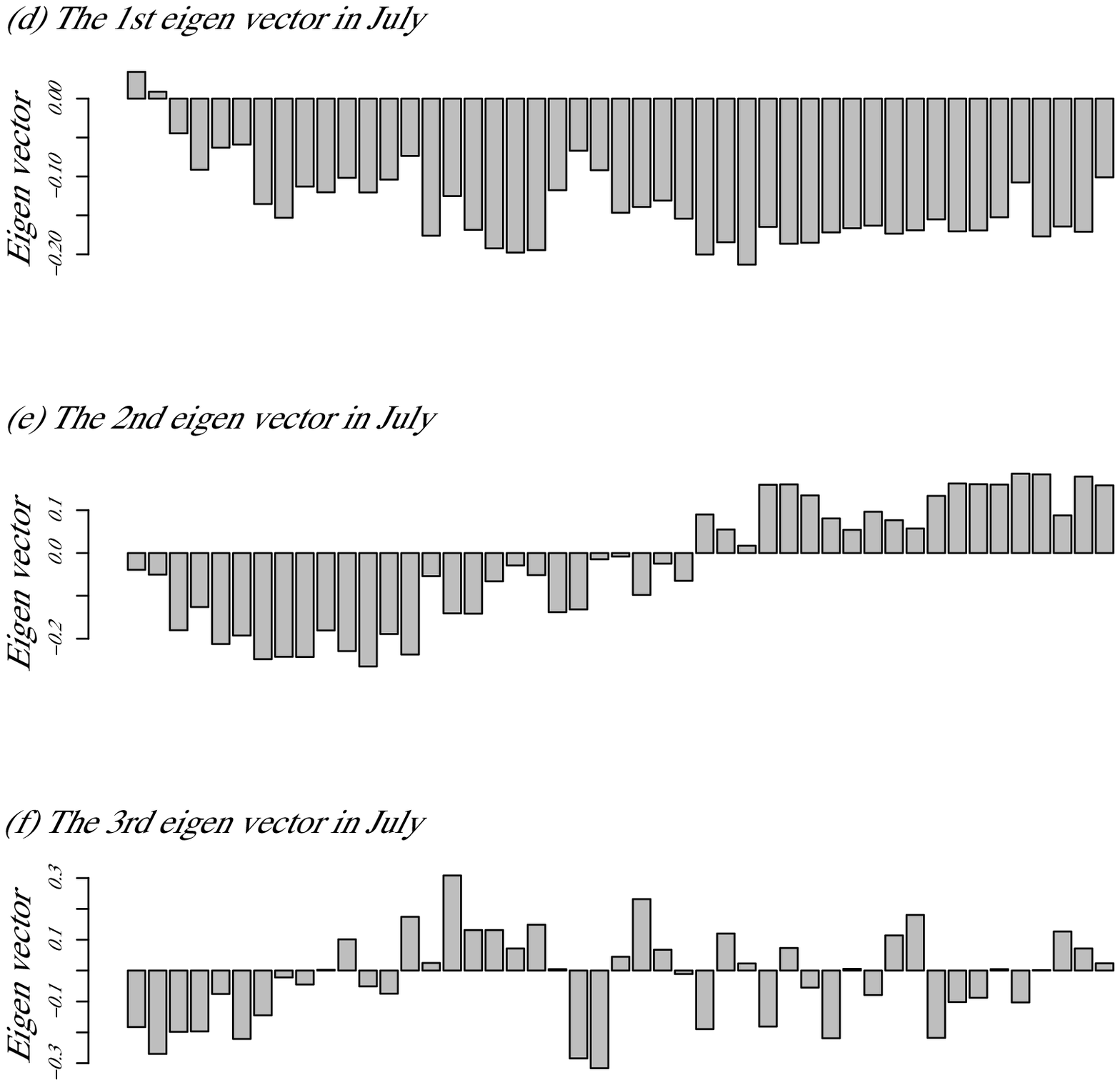}
\caption{
The 1st to 3rd eigen-vector for the whole of Japan in January and July
}
\label{fig:EigenVectorJapan}
\end{center}
\end{figure}

The cross-correlation coefficients for the whole of Japan are shown in Fig. \ref{fig:CrossCorrJapan}.
Panels (a) and (c) are genuine correlation $C^t$ and panels (b) and (d) are the random components $C^r$.
The genuine correlation $C^t$ was calculated using only the five largest eigen-values and corresponding eigen-vectors.
Figure \ref{fig:CrossCorrJapan} depicts that the genuine correlation $C^t$ has positive correlation and on the other hand the random components $C^r$ distributes around $0.0$.
The cross-correlation of PV output fluctuation in the Tokyo area was larger than the cross-correlation in the whole of Japan throughout the year. 
The 1st to 3rd eigen-vectors for the whole of Japan are shown in Fig. \ref{fig:EigenVectorJapan}. 
Forty seven components corresponds to all prefectures from Hokkaido to Okinawa in the whole of Japan.
The 1st eigen-vector has all components with the same sign. This means that the PV output fluctuates simultaneously in the same direction for all prefectures in the whole of Japan. The characteristic of the 1st eigen-vector in the whole of Japan is similar to the Tokyo area
The 2nd eigen-vector shows the weather change between eastern and western Japan.
The 3rd eigen-vector is more complicated. 
These characteristics of the 2nd to 5th eigen-vectors correspond to the smaller correlation coefficients in the whole of Japan.
However, it is noted that the coefficient of variation of PV output does not decrease proportionally to $N^{-1/2}$ as the number of PV sites N increased due to the observed cross-correlation among the sites. Thus, the so-called smoothing effect is expected to be smaller compared with the ideal case without cross-correlation. 


\subsection{Estimation of Forecast Error}\label{sec:cross:error}

We estimated the lower limit of the system-wide forecast error using the cross-correlation coefficients of the output fluctuation described in the previous section. 
Recently, numerical weather forecasting has gained higher accuracy, due to meteorological informations acquired by weather radars and meteorological satellites and the advancement of high performance computers.
Ultimately, as the forecast accuracy becomes higher, the forecast of the PV output time-series at each site converges on the moving average trend of the site. Thus, we expect that the short-term fluctuation will be the main component of the forecast error, because short-term fluctuation cannot be forecasted.
Therefore, we assume here that the lower limit of the forecast error is identical to the short-term fluctuation.

If the number of forecast sites is small, e.g., just one site in each prefecture, the system-wide forecast error involves the cross-correlation between the sites and consequently the system-wide error becomes large.
On the other hand, if the number of the forecast sites is large, the system-wide forecast error does not involve the cross-correlation between the sites and consequently the system-wide error becomes small.
If we consider that in near future installed PV systems are widely distributed in various places, the actual system-wide forecast error is expected to be between the above two extreme cases. 

We estimated the lower limit of the system-wide forecast errors and the coefficients of variation with/without considering the cross-correlations of the PV output fluctuation using Eqs. (\ref{eq:cross3}) and (\ref{eq:cross4}) with the genuine cross-correlation coefficient $\rho_{ij}$ shown in Figs. \ref{fig:CrossCorrTEPCO} and \ref{fig:CrossCorrJapan}.
The installed capacity of PV systems in 2030 was estimated by dividing the 100 GW capacity in the whole of Japan \cite{PVoutlook2030} proportionally to the demand of each prefecture. 
The estimations of errors and variation coefficients in the Tokyo area and the whole of Japan are shown in Tables \ref{ForecastErrorsTEPCO} and \ref{ForecastErrorsJapan}, respectively.
The 2nd to 5th columns of the tables represent error without correlation, coefficient of variation without correlation, error with correlation, and coefficient of variation with correlation, respectively. 
Both the system-wide forecast errors and the coefficients of variation are increased by considering the cross-correlation of the fluctuation.
The lower limit of the coefficients of variation in the Tokyo area is larger than the lower limit of the coefficients in the whole of Japan throughout the year. 

\begin{table}[!t]
\renewcommand{\arraystretch}{1.0}
\caption{Lower Limit of the System-Wide Forecast Errors in the Tokyo Area}
\label{ForecastErrorsTEPCO}
\centering
\begin{tabular}{c|c|c|c|c}
\hline
Month & Error w/o cor & Var w/o cor & Error w cor & Var w cor \\
\hline
\hline
Jan & $101.26$ & $0.0168$ & $181.36$ & $0.0302$ \\
Feb & $107.53$ & $0.0167$ & $206.78$ & $0.0321$ \\
Mar & $145.89$ & $0.0208$ & $296.55$ & $0.0423$ \\
Apr & $132.21$ & $0.0195$ & $246.95$ & $0.0365$ \\
May & $136.88$ & $0.0194$ & $262.23$ & $0.0373$ \\
Jun & $128.02$ & $0.0226$ & $221.59$ & $0.0391$ \\
Jul & $136.47$ & $0.0222$ & $256.34$ & $0.0417$ \\
Aug & $128.47$ & $0.0187$ & $236.19$ & $0.0344$ \\
Sep & $120.06$ & $0.0202$ & $239.98$ & $0.0405$ \\
Oct & $103.20$ & $0.0191$ & $183.09$ & $0.0338$ \\
Nov & $108.15$ & $0.0222$ & $236.34$ & $0.0485$ \\
Dec & $77.499$ & $0.0144$ & $154.74$ & $0.0288$ \\
\hline
\end{tabular}
\end{table}
\begin{table}[!t]
\renewcommand{\arraystretch}{1.0}
\caption{Lower Limit of the System-Wide Forecast Errors in the Whole of Japan}
\label{ForecastErrorsJapan}
\centering
\begin{tabular}{c|c|c|c|c}
\hline
Month & Error w/o cor & Var w/o cor & Error w cor & Var w cor \\
\hline
\hline
Jan & $148.36$ & $0.0093$ & $352.85$ & $0.0223$ \\
Feb & $164.13$ & $0.0090$ & $408.97$ & $0.0225$ \\
Mar & $218.45$ & $0.0103$ & $655.99$ & $0.0309$ \\
Apr & $207.07$ & $0.0094$ & $532.03$ & $0.0242$ \\
May & $202.87$ & $0.0089$ & $579.35$ & $0.0255$ \\
Jun & $196.43$ & $0.0108$ & $449.46$ & $0.0247$ \\
Jul & $205.63$ & $0.0104$ & $567.38$ & $0.0288$ \\
Aug & $191.47$ & $0.0087$ & $532.09$ & $0.0244$ \\
Sep & $184.43$ & $0.0099$ & $557.29$ & $0.0299$ \\
Oct & $160.52$ & $0.0092$ & $409.99$ & $0.0235$ \\
Nov & $165.50$ & $0.0113$ & $558.46$ & $0.0383$ \\
Dec & $124.18$ & $0.0083$ & $313.99$ & $0.0210$ \\
\hline
\end{tabular}
\end{table}

\section{Cost Estimation for PV Integration}\label{sec:cost}

\subsection{Unit Commitment Model}\label{sec:cost:model}

The purpose of our unit commitment model \cite{Ikeda2012, Ikeda2013} was to plan the operation schedule of thermal power plants so as to maximize the profit of an electric power utility by taking into account both the forecast of output and its error for renewable energies and the demand response of consumers on the change of electricity prices.
The essence of the model is described briefly as follows. 

\subsubsection{Objective Function}

The time series of the operational state of thermal power plant $i (i=1,\cdots,N)$ is obtained by maximizing the objective function: 
\begin{equation}
\begin{split}
F(p_t^i,u_t^i,z_t^i,w_t^l) = & \sum_{t=1}^T d_t^{(f)} \sum_{l=1}^{L} w_t^l r^l \Big( \frac{r^l}{\bar{r}} \Big)^{\epsilon_d} \\ 
& - \sum_{t=1}^T \sum_{i=1}^N [b_i p_t^i + S_i z_t^i].
\label{eq:ObjFnc}
\end{split}
\end{equation}
This objective function represents the profit of an electric power utility.
The first term of the r.h.s. in Eq. (\ref{eq:ObjFnc}) is the sales revenue and the second term is the operation cost.
Here $N$, $T$, and $L$ are the number of thermal power plants, time horizon, and number of price levels, respectively.
Continuous variables $p_t^i$ is the output power variable of thermal power plant $i$, and integer variables $u_t^i$, $z_t^i$, and $w_t^l$ are	
the status production variable of thermal power plant $i$ (1=committed, 0=decommitted),
start-up variable of thermal power plant $i$ (1=start up, 0=others), and 
demand response variable (1=selected, 0=not selected), respectively.
Parameters $S_i$ and $b_i$ represent the start-up cost of thermal power plant $i$ and the fuel cost of the thermal power plant $i$, respectively. 
The forecasted demand and its error are indicated by $d_t^{(f)}$ and $\sigma_d$, respectively. Here, $(f)$ stands for forecasting.

\begin{figure}
\begin{center}
\includegraphics[scale=0.5]{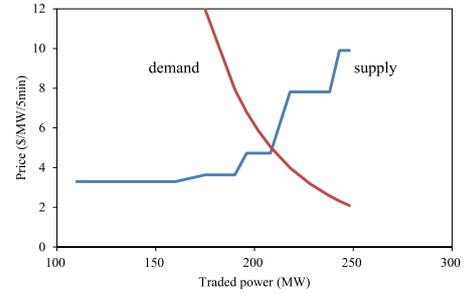}
\caption{Demand and supply}
\label{fig:ds}
\end{center}
\end{figure}

Other parameters related to the demand response $\bar{r}$, $r^l$, $\epsilon_d$ are the average electricity price, price level, and price elasticity of demand, respectively. 
If the electricity price $r$ deviates from the average price $\bar{r}$, the demand $d$ is changed from the average demand $\bar{d}$ as follows:
\begin{equation}
\frac{d}{\bar{d}} = \Big( \frac{r}{\bar{r}} \Big)^{\epsilon_d}.
\label{eq:Dcurve}
\end{equation}
The dependence of demand $d$ on price $r$ is depicted in Fig. \ref{fig:ds}.

\subsubsection{Global Constraints}

The sum of the demand response variable $w_t^l$ has to satisfy the constraint
\begin{equation}
\sum_{l=1}^{L} w_t^l = 1
\label{eq:Plevel}
\end{equation}
to ensure that only a single price level $r^l$ is selected.
In addition to this constraint, the average of the selected price $r^l$ has to be equal to the average price $\bar{r}$
\begin{equation}
\frac{1}{T} \sum_{t=1}^T \sum_{l=1}^{L} w_t^l r^l \le \bar{r}.
\label{eq:meanP}
\end{equation}
Moreover the total demand has to be unchanged by the demand response:
\begin{equation}
\sum_{t=1}^T d_t^{(f)} = \sum_{t=1}^T \widetilde{d_t^{(f)}}, 
\label{eq:Dconserve1}
\end{equation}
\begin{equation}
\widetilde{d_t^{(f)}} = d_t^{(f)} \sum_{l=1}^{L} w_t^l \Big( \frac{r^l}{\bar{r}} \Big)^{\epsilon_d}. 
\label{eq:Dconserve2}
\end{equation}
Importantly, the sum of supply has to be greater than the demand:
\begin{equation}
\sum_{i=1}^N p_t^i+wd_t^{(f)}+pv_t^{(f)} + g_t - h_t \ge \widetilde{d_t^{(f)}}.
\label{eq:DSbalanceNoSigma1}
\end{equation}
where $wd_t^{(f)}$, $pv_t^{(f)}$, $g_t$, and $h_t$ are the forecasted wind power generation, forecasted PV generation, discharged power from pumped hydro power, and charging to pumped hydro power, respectively.

If we consider the forecast error of demand $\sigma_d$, forecast error of wind power $\sigma_w$, and forecast error of PV $\sigma_p$,
the constraint in Eq.(\ref{eq:DSbalanceNoSigma1}) can be rewritten as
\begin{equation}
\frac{\sum_{i=1}^N p_t^i+wd_t^{(f)} + pv_t^{(f)} + g_t - h_t - \widetilde{d_t^{(f)}}}{\sqrt{\sigma_d^2+\sigma_w^2+\sigma_p^2}} \ge \phi^{-1}(\alpha),
\label{eq:DSbalance}
\end{equation}
where $\alpha$ and $\phi(\cdot)$ are the probability to ensure the supply-demand balance and the cumulative distribution function, respectively.
In this paper we assumed that the system-wide error distribution is the normal distribution.

Pumped hydro power has to satisfy the constraints:
\begin{equation}
v_t c^{min} \le g_t \le v_t c^{max},
\label{eq:pumpedhidro1}
\end{equation}
\begin{equation}
(1-v_t) c^{min} \le h_t \le (1-v_t) c^{max},
\label{eq:pumpedhidro2}
\end{equation}
\begin{equation}
R^{min} \le \sum_{s=1}^t (h_s \eta - g_s) \Delta_t \le R^{max},
\label{eq:pumpedhidro3}
\end{equation}
where $v_t$, $c^{min}$, $c^{max}$, $R^{min}$, $R^{max}$, $\eta$, and $\Delta_t$ are the state variable of the pumped hydro power (1=discharge, 0=charge), minimum discharge power, maximum discharge power, minimum stored energy, maximum stored energy, efficiency, and time step, respectively.

\subsubsection{Local Constraints for Thermal Power Plants}

The following constraints are used for each thermal power plant as typical constraints in a unit commitment model.

\begin{itemize}
\item Generation Capacity

The output power $p_t^i$ has to be between the maximum output power $\bar{p}_{max}^i$ and the minimum output power $\bar{p}_{min}^i$ when the operation is in steady state:
\begin{equation}
\bar{p}_{min}^i u_t^i \le p_t^i \le \bar{p}_{max}^i u_t^i.
\label{eq:Capa}
\end{equation}

\item Ramp-up Limit

The increase in the output of thermal power plant $i$ should be smaller than the maximum ramp-up speed $\Delta_+$ when the unit is up at the previous time step
and is smaller than the minimum output power $\bar{p}_{min}^i$ when the unit is down at the previous time step:
\begin{equation}
p_t^i - p_{t-1}^i \le u_{t-1}^i \Delta_+^i + (1-u_{t-1}^i) \bar{p}_{min}^i.
\label{eq:RumpUp}
\end{equation}

\item Ramp-down Limit

The decrease in the output of thermal power plant $i$ should be smaller than the maximum ramp-down speed $\Delta_-$ when the unit is up at time step $t$
and is smaller than the maximum output power $\bar{p}_{max}^i$ when the unit is down at time step $t$:
\begin{equation}
p_t^i - p_{t-1}^i \ge - u_t^i \Delta_-^i - (1-u_t^i) \bar{p}_{max}^i.
\label{eq:RumpDown}
\end{equation}

\item Minimum Up-time Constraint

Thermal power plant $i$ has to be operated longer than the minimum up-time requirement $\tau_+^i$, once the unit is up:
\begin{gather}
u_t^i \ge u_s^i - u_{s-1}^i, \notag \\
s \in [t-\tau_+^i, t-1].  
\label{eq:MinUp}
\end{gather}

\item Minimum Down-time Constraint

Thermal power plant $i$ has to be stopped longer than the minimum down-time requirement $\tau_-^i$, once the unit is down:
\begin{gather}
u_t^i \le 1 + u_s^i - u_{s-1}^i, \notag \\
s \in [t-\tau_-^i, t-1]. 
\label{eq:MinDown}
\end{gather}

\item Constraint on the Start-up Variable

The start-up variable $z_t^i$ has to satisfy the following constraints by definition:
\begin{gather}
z_1^i \ge u_1^i,  \notag \\
z_t^i \ge u_t^i - u_{t-1}^i (t>2).
\label{eq:Start}
\end{gather}

\end{itemize}

\subsection{Integration Cost}\label{sec:cost:estimation}

\begin{figure}
\begin{center}
\includegraphics[width=0.45\textwidth]{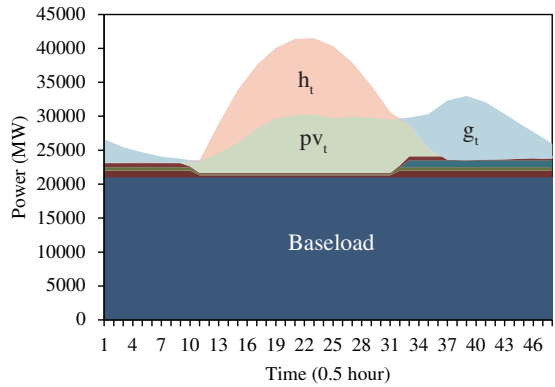}
\caption{
Power from various power plants to satisfy the demand in the early May
}
\label{fig:ShareOfPlants}
\end{center}
\end{figure}

We estimated the effect of increase of the forecast error on the operation cost of thermal power plants by using the unit commitment calculation for the Tokyo area.
In the unit commitment calculation, we used the estimated system-wide forecast error shown in Tables \ref{ForecastErrorsTEPCO} and \ref{ForecastErrorsJapan} and the model parameters shown in Table \ref{SystemParameters}.
In Fig. \ref{fig:ShareOfPlants}, power from various power plants to satisfy the demand in early May is shown for the Tokyo area. We had the smallest demand and largest output from PV systems in this season. Therefore, the condition for PV integration is toughest throughout the year. 
Fig. \ref{fig:ShareOfPlants} depicts that many thermal power plants stand by for the PV output fluctuation.

The integration cost $\epsilon$ of the PV system per unit output energy was estimated by
\begin{equation}
\epsilon = \frac{C(\sigma_p)-C(\sigma_p=0)}{\sum_{t=1}^T pv_t \Delta_t},
\label{eq:IntegCost}
\end{equation}
where $C(\sigma_p)$ and $C(\sigma_p=0)$ are the operation cost with forecast error and operation cost without forecast error, respectively.
The main component of the integration cost $\epsilon$ is due to the balancing capability.

\begin{figure}
\begin{center}
\includegraphics[width=0.45\textwidth]{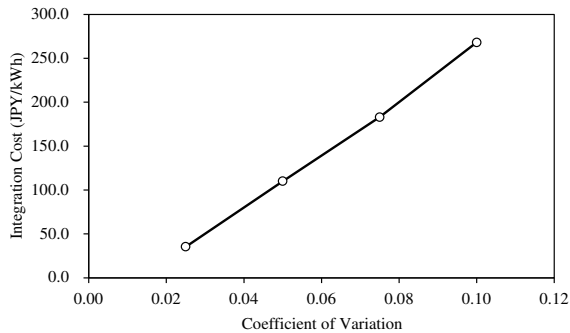}
\caption{
Integration cost as a function of the forecast error
}
\label{fig:IntegCost}
\end{center}
\end{figure}

The estimated integration cost is shown in Fig. \ref{fig:IntegCost} as a function of the forecast error. 
It is noted that the integration cost is comparable to the average electricity price $\bar{r}$ at the lower limit of the coefficient of variation as shown in Table \ref{ForecastErrorsTEPCO}.
It is evident that the integration cost per unit output energy is the smaller for the whole of Japan due to the small coefficient of variation.

\begin{table}[!t]
\renewcommand{\arraystretch}{0.9}
\caption{Parameters used in the unit commitment calculation}
\label{SystemParameters}
\centering
\begin{tabular}{c|c}
\hline
symbol & parameters \\
\hline
\hline
$N$ & 91 \\
$T$ & 48 \\
$\Delta_t$ & 30 min. \\
$L$ & 20 \\
$\bar{r}$ & 30 JPY/kWh \\
$r^1$ & 20 JPY/kWh \\
$r^L$ & 40 JPY/kWh \\
$\epsilon_d$ & 0 (without DR), -0.1 (with DR) \\
$\alpha$ & 1.28 (90 $\%$ CL for the normal distribution) \\
$\sigma_w$ & 10$\%$ of the wind output power \\
$\sigma_d$ & 5$\%$ of the system load \\
$c^{min}$ & 0.0 MW \\
$c^{max}$ & 11800.0 MW \\
$R^{min}$ & 0.0 MWh \\
$R^{max}$ & 118000.0 MWh \\
$\eta$ & 0.7 \\
baseload & 21000.0 MW \\
\hline
\end{tabular}
\end{table}

\section{Discussion}\label{sec:discussion}

The validity of the concept of ``local production for local consumption of renewable energy" and alternative policy implications are discussed.

When the installed capacity of the PV system is small, the ``local production for local consumption of renewable energy" is economically feasible. 
The area price in the Tokyo area could be high enough to be close to the feed-in tariff price for PV power and transmission loss is mitigated due to limited transmission inside the Tokyo area.
According to the installation plan, the capacity of PV system will become large, parallel with the vitalization of the power market. In this phase, the lack of balancing capability becomes obvious and the integration cost exceeds the current electricity price.
Consequently, the ``local production for local consumption of renewable energy" concept becomes infeasible.

In the near future, we will expand the capacity of the inter-connections between the Hokkaido and Tohoku area, the Tohoku and Tokyo area, and the 50Hz/60Hz boundary in the west of the Tokyo area. A new role is expected for the transmission systems. That is the reduction of the coefficient of variation of the PV output. For instance, the development of the high-voltage direct current (HVDC) line in the whole of Japan \cite{Tanaka2012} will reduce the requirement for the balancing capability. A novel methodology is desired in order to plan an optimal transmission system.

\section{Conclusion}\label{sec:conclusion}

We analyzed the cross-correlation of PV output fluctuation for the actual PV output time series data in both the Tokyo area and the whole of Japan using the principal component analysis with the random matrix theory. Based on the obtained cross-correlation coefficients, the lower limit of the forecast error for PV output was estimated with/without considering the cross-correlations. Both the system-wide forecast errors and the coefficients of variation were increased by considering the cross-correlation of the fluctuation. The lower limit of the coefficients of variation in the Tokyo area was larger compared with that of the whole of Japan throughout the year. 
Then, the operation schedules of thermal plants were calculated to integrate PV output using our unit commitment model with the estimated forecast errors. The integration cost of PV system was also estimated. The integration cost was comparable to the average electricity price $\bar{r}$ (30JPY/kWh) at the lower limit of the coefficient of variation. Finally, validity of the concept of ``local production for local consumption of renewable energy" and alternative policy implications were discussed. It was evident that the integration cost per unit output energy is smaller for the whole of Japan due to the small coefficient of variation. This means that the concept of ``local production for local consumption of renewable energy" is not economically feasible. The development of the transmission lines planned in the near future will reduce the balancing capability required for PV integration.

% conference papers do not normally have an appendix

% use section* for acknowledgement
%\section*{Acknowledgment}
%The authors would like to thank...

% trigger a \newpage just before the given reference
% number - used to balance the columns on the last page
% adjust value as needed - may need to be readjusted if
% the document is modified later
%\IEEEtriggeratref{8}
% The "triggered" command can be changed if desired:
%\IEEEtriggercmd{\enlargethispage{-5in}}

% references section

% can use a bibliography generated by BibTeX as a .bbl file
% BibTeX documentation can be easily obtained at:
% http://www.ctan.org/tex-archive/biblio/bibtex/contrib/doc/
% The IEEEtran BibTeX style support page is at:
% http://www.michaelshell.org/tex/ieeetran/bibtex/
%\bibliographystyle{IEEEtran}
% argument is your BibTeX string definitions and bibliography database(s)
%\bibliography{IEEEabrv,../bib/paper}
%
% <OR> manually copy in the resultant .bbl file
% set second argument of \begin to the number of references
% (used to reserve space for the reference number labels box)

% that's all folks
\end{document}